# Multiple Criteria Decision-Making Preprocessing Using Data Mining Tools

A. Mosavi

**Faculty of Informatics, University of Debrecen**
**Debrecen 4032 , Hungary**

**Abstract**
Real-life engineering optimization problems need Multiobjective Optimization (MOO) tools. These problems are highly non-linear. As the process of Multiple Criteria Decision-Making (MCDM) is much expanded most MOO problems in different disciplines can be classified on the basis of it. Thus MCDM methods have gained wide popularity in different sciences and applications. Meanwhile the increasing number of involved components, variables, parameters, constraints and objectives in the process, has made the process very complicated. However the new generation of MOO tools has made the optimization process more automated, but still initializing the process and setting the initial value of simulation tools and also identifying the effective input variables and objectives in order to reach the smaller design space are still complicated. In this situation adding a preprocessing step into the MCDM procedure could make a huge difference in terms of organizing the input variables according to their effects on the optimization objectives of the system. The aim of this paper is to introduce the classification task of data mining as an effective option for identifying the most effective variables of the MCDM systems. To evaluate the effectiveness of the proposed method an example has been given for 3D wing design.

## 1. Introduction

Engineering optimization plays a significant role in today's design cycle and decision-making. The optimization process is essentially seen as system improvement in order to identify and arrange the effective variables. Problems with multiple objectives and in multiple disciplines are known as Multiobjective optimization decision-making or MCDM problems. However, many real-life phenomena have a nonlinear nature. Therefore nonlinear tools for handling several objectives are needed. In this situation nonlinear MOO deliver an extensive, self-contained approach. [2]

Nonlinear MOO means MCDM dealing with nonlinear functions of decision variables. Identification of the optimum solution of a nonlinear multiobjective problem and decision-making is often not possible because of the size of the problem and lack of knowledge about effective variables. [2]

Decision-making in the problems related to more than one objective, originating in several disciplines, has been a challenge to scientists for a long time. They have been asked to solve problems with several conflicting objective functions. Basically using a single objective optimization technology is not sufficient to deal with real-life engineering optimization problems. MOO tools generate the solutions which are called Pareto frontier solutions and the final decision could be one of these solutions.

Generating Pareto frontier solutions and following it engineering MOO decision-making process has become automatic and made easier by utilizing special packages developed based on different MOO algorithms and techniques such as modeFRONTIER and IOSO families. But still the whole process needs improvements as the mentioned packages always have limitations in handling the large amounts of data such as input variables, constraints, objectives and visualization. However increasing the number of input variables which is the characteristic of MCDM engineering systems has made the process more complicated. The decision-maker setting up a new project and making decisions faces a high amount of variables and objectives. In this regard utilizing new alternative techniques of data mining in place of traditional expert-based methods in different areas of MCDM including MOO, visualization and decision-making appeared to be very popular and supportive in dealing with engineering optimization problems which have left lots of room for researches.

## 2. Applications of Data Mining in MCDM process

MCDM consists of two parts, MOO and Multiple Criteria Decision Analysis (MCDA). The involved data set in both MCDM branches are likely to be huge and complex. Large-scale data of MCDM problems can only be handled with the aid of computers. The field of knowledge discovery, or data mining, has evolved in the recent past to address the problem of automatic analysis of larger amounts of data. However, processing commands may need to be entered manually by data analysts and data mining results can be fully used by decision makers only when the results are understood explicitly.

Different tasks of data mining including description, estimation, prediction, classification, clustering and, association have been utilized in different applications of MCDM [13, 24, 25, 22, 23].

Data reduction is an essential purpose in the data mining process. Data reduction is developed to fulfill objectives such as improving accuracy of models, scaling the data mining models, reducing computational cost, and providing a better understanding of data. The aim of data reduction is to find a subset of attributes which represent the concept of data without losing important information.

Data mining techniques in the applications of MCDM are applicable to the data sets which are needed to be analyzed before the MOO process as a preprocessing sequence and also to the Pareto fronts solutions and contributions of design variables in decision-making process. Hence the applications of data mining in MCDM could be separately studied in three different parts of MOO, decision-Making and visualization. In the next three sessions these applications are reviewed.

### 2.1 MOO and Data Mining

Large quantities of approaches have been developed for solving nonlinear MOO problems. A classification of these methods has been suggested in [31].In some of these methods the data mining tools have been applied in order to make the process easier, minimize the computational cost, etc. for instance Zitzler et al. [22] have integrated different MOO techniques and applied the clustering task of data mining called "average linkage method" [23] to maintain diversity.

### 2.2 Visualization and Data Mining

Graphs and plots are usually applied for understanding maximum three-dimensional relationships achieved between MOO objectives. But for visualization the multiple objective problems data mining tasks have to be applied. In this regard Classifications and Clustering [26, 27] are the most useful tasks. Common data mining methods utilized for classification are the *k*-nearest neighbor decision tree, and neural network [28]. Obayashi et al. [27] utilizes the clustering technique of data mining for visualizing the four objectives of optimization in a self-organizing map. Without the aid of data mining the visualization of the huge amounts of data in MOO is extremely difficult. For instance dealing with the computational complexity of heatmap-based MOO visualization in [29] is completely dependent on the clustering method.

### 2.3 Decision-Making and Data Mining

Decision making is a post-processing tool which helps the user to make selection of the best designs from a family of Pareto solutions. Fig. 2 shows the process of decision making where a decision maker selects one course of action from many possibilities.

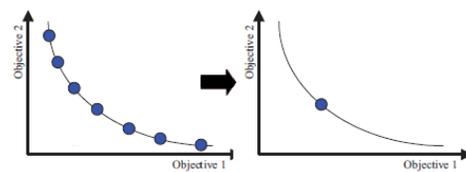

Fig.1 Pareto-optimal solutions and decision-making

The MOO approach attempts to find the Pareto solutions set, although the decision-making procedures are also very important to choose one particular solution from the Pareto solutions set for implementation in MCDM process.

In dealing with MCDM problems, the final obtained solution must be as close to the true optimal solution as possible and that solution must satisfy the supplied preference information. In dealing with such a task, input data to MCDM such as initial value of variables is extremely important. An additional difficulty is the fact that the decision maker or optimizer is not necessarily an expert in the field of the decision making process so as to be able to correctly identify effective and valuable variables. Hence, getting help for analyzing the input variables and decision-making variables from an intelligent computational system seems to be necessary. For instance Katharina et al [13] utilized a data mining tool for supporting the process of decision making. In MCDM, Satisfying Trade-Off Method (STOM) is a tool for dealing with decision making problems. Nakayama [24, 25] in some multiobjective STOM problems utilizes the classification task of data mining to satisfy the decision making procedure.

## 3. Meta-modeling Based Multiobjective Optimization Tasks

Meta-modeling or surrogate modeling tasks are applied in order to model the design space on the basis of the limited number of numerical analyses. Meta-modeling based MOO tasks have wide application in engineering MCDM problems. In these problems there are several sources of complexity, such as the computational difficulties in modeling, and the high number of variables, objectives and constraints. Besides, the coupling process of different disciplines is a challenging job. For reasons of simulation, there are software packages which are integrated into the workflow. A Limited number of simulations could be run in the limited period of time. In this situation Meta-modeling based tasks of MOO are utilized for obtaining maximum information from a minimum number of simulations. Fig. 1 shows the general workflow of Meta-modeling based MOO.

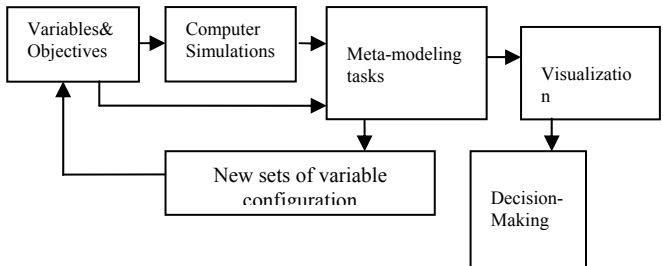

Fig.2 General workflow of Meta-modeling based MOO

In this workflow the variables are identified and initialized at the first step. Then they pass directly to the next steps of numerical analysis and MOO. Therefore, there is not any control and monitoring of the input variables. "*Compressor blade optimization*" by Xinwei et al. [3] is an example of this workflow.

For dealing with MCDM problems in order to obtain the Pareto frontier solutions and, following it, the decision-making process, one or more of the Meta-modeling tasks below have been utilized depending on the characteristics of the problem.

- Design of Experiments (DOE)
- Genetic algorithms(GA)
- Response Surface modeling(RSM)
- Hybrid system
- Indirect Optimization on the basis of Self-Organization (IOSO)

3.1 Modeling; State of the Problem

Before any optimization can be done, the problem must be modeled first. In this regard identifying all dimensions of the problem such as formulation of the optimization problem with specifying input variables, decision variables, objectives, constraints, and variable bounds is an important task [19]. However in some cases the problem is not clear in terms of input variables. The proposed method of this paper tries to find the variables which have greater effects on objective functions. This would support the MCDM processes in uncertain sampled records in order to estimate the whole design space.

The approach is to mine the problem's data set including input variables and their effects on objectives. It is supposed to help with gaining a better understanding of the design space.

## 4. The Effects of Large Amounts of Input Variables on MOO Tasks; Research Motivation

On one hand increasing the number of input variables in a MOO problem by applying the Meta-modeling tools would increase the accuracy of the results. On the other hand it can also cause computational difficulty and makes the process very complicated. Identifying the input variables of a design optimization problem is a manual step of initializing the optimization workflow. This step is usually done by an engineer and he is responsible for adding or ignoring certain variables based on his understanding of the system. In the sections below the difficulties of dealing with a large amount of input variables in Meta-modeling based MOO tasks are reviewed. Dealing with these difficulties has motivated this research.

4.1 DOE

DOE explores the design space and automatically chooses the minimum set of solutions which contain the maximum amount of information. This ability helps to achieve smaller design space. Each single numerical analysis takes a long time and some simulations are very expensive to run for more than a limited number of calculations. In this regard DOE delivers enough initial calculations which allow the optimization process to learn the behavior of design parameters. DOE mostly deals with the value of variables, variable variations and the properties of the governing parameters. However ranking, organizing and removing the less important variables as a supporting process, before setting up the DOE projects, could be very useful in reducing the design space and computational costs.

In business systems engineering, Vergidis et al. in "*Optimization of business process designs*" [9] and Laszlo et al. in the field of management and resolution in "*resource allocation under uncertainty*" [10] use DOE tools for MOO. In their analyzing processes or other related approaches experiments have been used to evaluate which inputs have more impact on the outputs, and what the target level of tested inputs might achieve on a desired output. However in large design space where lots of input variables are involved having experiments in all design space results in excessive computational costs. In this situation data mining classification techniques can reduce the computational costs as it could deliver useful information about those parts of design space where there is not a great deal of experimental data available. In both the above mentioned approaches input variables are in high number which may not have any effect on system behavior but still were included in the workflow. This fact has increased the computation time of DOE. Classification of the input data sets of the DOE process could deliver more information about design space to minimize the computational costs.

4.2 GA-based MOO

GA is very important in the developing of optimization techniques. The GA tools work with random variables. The GA-based algorithm searches the design space according to the different variable configurations while utilizing an objective or fitness function. GA approaches have been widely utilized for shape optimization such as the application presented by Arularasan et al [11]. In most of the GA applications in multiobjective shape optimization problems there are more than fifty design variables involved, including the variation of constraints. It has been observed that such processes involve dealing with a huge amount of data. As a result minimizing the number of variables in the GA-based shape optimization process, even in the same amount of constraints, could make the optimization design space smaller and therefore dramatically decrease the computational cost. Zhong et al. in "*Robust Airfoil Optimization*" [4] and András et al. "*in Aerodynamic optimization*" [5] utilized genetic algorithm-based methods for aerodynamic Multiobjective shape optimization. In their study managing the number of input variables is introduced as an issue. Hence, it is assumed that removing the less effective variables from the GA-based design space and searching with a minimum number of variables to find the optimal solutions, will make the process less complicated.

In the other fields such as power systems and modeling of economic process; Li in "*Study of multi-objective optimization and multi-attribute decision-making*" [8] utilizes genetic algorithm tools in the field of marine systems with an emphasis on safety issues and Alan et al. in "*Optimization of crashworthy marine structures*" [7] has applied GA and response surface tools together for optimization. In above mentioned examples of large scale applications the GA faces a huge number of variable configurations, which is an expensive computation process. In dealing with such large scale problems an automated preprocessing tool needs to be applied to identify the most important variables out of thousands possible variables which have more effects on optimization objectives.

4.3 Response Surface

In cases where running a full optimization is not practical virtual optimizations have become effective by utilizing the response surface tools. There is lots of theoretical and practical literature available in this regard [7, 19]. In RSM-based MOO creating the Meta-models based solely on the most important variables rather than all variables, leads to more accurate models of estimation.

4.4 Hybrid systems

Different tasks of Meta-modeling MOO are combined to obtain some hybrid methods. A hybrid method tries to exploit the specific advantages of different approaches by combining more than one together. Hybrid methods are a combination of important group of methods that have significantly contributed to the renewal of MOO process. For example, the modeFRONTIER package has combined the robustness of the GA with the accuracy of gradient-based methods. But still these packages need some support in order to make the design space as small as possible. The preprocessing of input values could make a huge difference in terms of computational costs. Hybrid-based tools and packages are very powerful for dealing with MOO problems. In the field of intelligent design and manufacturing systems, Olcer in "*A hybrid approach for multi-objective combinatorial optimization problems in ship design"* [10] utilized hybrid tools. In his approach it seems that ranking the input variables based on their effects on the design objectives before utilizing any kind of Hybrid-based approaches could be very efficient.

4.5 Indirect Optimization on the basis of Self-Organization (IOSO)

IOSO technology is based on the response surface technology. Its technique is totally different from the current approaches to optimization. The applied strategy has higher efficiency and provides wider range of capabilities than standard algorithms. The main advantage of IOSO technology is ability to solve very complex optimization tasks. It can approximate objective functions with complex topology using minimal number of points in the experiment plan, particularly including the case when the number of points is less than the number of design variables [30, 21]. However because of the limitation of the package in accepting a limited number of input variables the presented preprocessing method of this paper could work with IOSO to gain better results.

### 4.5.1 Case Study

In an approach to find the optimal Pareto solutions for the complex and nonlinear mathematical problem of *designing the curves by multicriteria optimization*, including three conflicting objectives, an IOSO-based technique has been utilized [21]. The aim was to incorporate several design objectives into a single optimization process. There was numerous numbers of input variables therefore the problem was modeled several times with a different number, and different configuration, of input variables. It was observed that including or excluding some variables from the same class could make a huge difference to the final estimated model. Therefore the preparation phase of selecting the variables should be done with complete accuracy. This fact has motivated current research to try alternative speedy techniques besides those of DOE tools- in order to manage the design space- by classifying, identifying, ranking and/or removing variables based on the training data sets.

## 5. Preprocessing and Data Mining

As was mentioned the different tasks of MOO and decision-making in engineering optimization applications applying Meta-modeling have the common difficulty of dealing with the large amounts of input variables, design variables, decision variables and objectives. The decision-maker often has no idea about the importance of the variables. Thus it is difficult to organize the number of variables based on expert knowledge. Additionally variable ranking is also a difficult task, especially when several computer simulations, objectives and decision makers are involved.

The decision maker in creating a new project faces a high amount of variables and objectives which makes the process very complex. Ranking or identifying the less important variables and objectives, and following it, reducing the number of variables and even objectives which have minimum effects on results, could make the process less complicated and speedier. In this regard, there haven't been adequate publications yet though it was assumed that analyzing the inputs and outputs of engineering numerical analysis for some records could deliver enough information for estimating the whole system behavior.

The most related work has been done by Obayashi et al [27]. They utilize the Analysis Of Variance (ANOVA) approach of data mining and the effect of each design variable to the objectives and the constraint functions in a quantitative way. ANOVA uses the variance of the model due to the design variables on the approximation function. By their proposed method, applying the data mining task of clustering, the effect of each design variable on the objective functions can be calculated and visualized.

It was proved that data mining tasks could be applied for processing the inputs' and outputs' data of numerical analysis systems. Fig.3 describes the expected data preprocessing step in the general workflow of MCDM process. The data set of problems including the numerical analysis records for some calculations before the MOO process in a preprocessing step are analyzed applying the classification of data mining. After preprocessing the design space is reduced which will make the rest of the process less complicated.

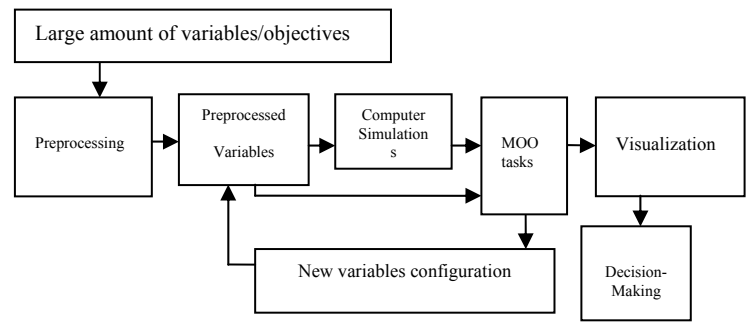

Fig.3 General Workflow for MOO including preprocessing

### 5.1 Proposed Methodology

Our proposed method [6] is based on the classification task of data mining showing the effect of each design variable on the objectives. In this method the *target categorical variable* is defined as the result value of numerical analysis performed by any of the Computer Aided Engineering packages. The target categorical variable might be partitioned in different classes. We would classify the target categorical variable for the experiments and their dependent input variables which are not in our database based on other characteristics associated with them. The classification algorithm would examine the data set containing both the input variables and the classified target variable. Then the algorithm would learn about which combinations of input variables are associated with which class of target categorical variable. The achieved knowledge will deliver the *training set*.

The data mining classifier package of Weka provides implementations of learning algorithms and data sets which could be preprocessed and feed into a learning scheme, analyzing the classifier results and its performance.

Note, that Weka includes methods for all the standard data mining problems such as regression and classification which are necessary for the proposed approach. Weka also includes many data visualization facilities and data preprocessing tools.

One algorithm of WEKA, the BFTree, is chosen to search the whole design space for the input variables where there are no records of target categorical variable. Based on the classifications in the training set, the algorithm would be able to classify these records as well.

As the computational simulations by most of the engineering packages are very expensive and time consuming, the data sets of most Meta-modeling based optimization problems are unable to find the information of the whole design space. In this situation classification could work efficiently to estimate the entire design space. The workflow of proposed methodology is described in Fig.4. Here the classification method is utilized to create several classifiers or decision trees. In the next steps the most important variables which have more effects on the objectives are selected.

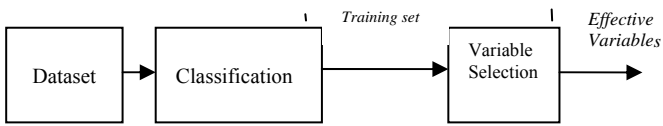

Fig.4 proposed methodology workflow

Regressions and model trees are constructed by a decision tree to build an initial tree. However, most decision tree algorithms choose the splitting attribute to maximize the information gain. It is appropriate for numeric prediction to minimize the intra subset variation in the class values under each branch. The splitting criterion is used to determine which variable is the better to split the portion T of the training data. Based on the treating of the standard deviation of the objective values in T, as a measure of the error and calculation, the expected reduction in error as a result of testing each variable is calculated. The variables which maximize the expected error reduction are chosen for splitting. The splitting process terminates when the objective values of the instances vary very slightly, that is, when their standard deviation has only a small fraction of the standard deviation of the original instance set. Splitting also terminates when just a few instances remain.

The Mean Absolute Error (MAE) and Root Mean Squared Error (RMSE) of the class probability is estimated and assigned by the algorithm output. The RMSE is the square root of the average quadratic loss and the MAE is calculated in a similar way using the absolute instead of the squared difference.

In the first preprocessing approach-utilizing the proposed method [6] - the database was created by twelve computational simulation runs, forty two variables and three objectives. Three different data mining classification algorithms were applied (J48, BFTree, LADTree) and their performance was compared in order to choose attribute importance. As the result LADTree was found to be a better choice for further classifications.

The intention in Meta-modeling based optimization problems due to costly computational simulation is to run minimum simulation runs as much as possible. Thus in the given example there was an attempt to minimize the number of simulations to five calculations.

## 6. Given Example

The example is given in aerospace engineering where the structural simulation is tightly integrated into more than one discipline and criterion. Meanwhile, the trend nowadays is to utilize independent computational codes for each discipline. In this situation, the aim of MCDM tools is to develop methods in order to guarantee that effective physical variables be involved. In order to approach the optimal shape in aerospace engineering optimization problems, the MOO techniques are asked to deal with all important objectives and variables efficiently. The example has been given in shape optimization of a 3D airfoil with objectives in displacements distribution. In the similar cases in [4, 5, and 14] for aerodynamic optimization of a 3D wing there was an attempt to utilize the MOO techniques applying GA in a multidisciplinary environment. In their approach the geometry of points X and Y are actually input variables. All possible variables have been involved in the optimization process ignoring their effects on objectives. It is assumed that the preprocessing of the available records of simulations can identify the most effective variables and running the MOO with just effective variables could dramatically decrease the MOO computational costs. The approaches such as applied MOO GA-based in [14] could be more effective and less complicated, taking less computation time, if they were modeled just by effective variables.

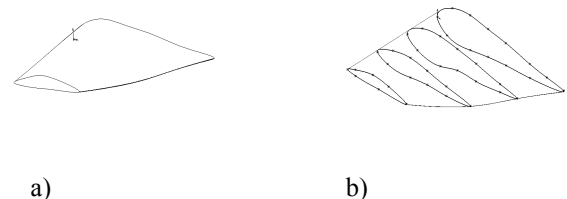

a)          b)

Fig.5 Airfoil geometry modeled by S-plines

The airfoil of Fig.5 part (a) is subjected to shape optimization. The shape needs to be optimized in order to deliver minimum displacement distribution in terms of applied pressure on the surface. Fig.5 (b) shows the basic curves of the surface modeled by S-plines. For modeling the surface four profiles have been utilized with forty two points. The coordinates of all points are supplied by a digitizer, each point includes three dimensions of X, Y, and Z. Consequently there are 126 columns plus two objectives. Objectives are listed as follow:

- Objective1(O1): Minimizing the displacements distribution in the airfoil for constant pressure value of α
- Objective2(O1): Minimizing the displacements distribution in the airfoil for constant pressure value of 2α

An optimal configuration of forty two variables is supposed to satisfy the two described objectives.
In the described MOO preprocessing the number of variables is subjected to minimization before any MOO process takes place in order to reduce the design space and computational costs.

Table 1:   data sets including five calculations' results

| | Variables Configuration : V1-V42 | CAD Model | Displacement Distribution | Objective Results |
|---|---|---|---|---|
| 1 | 0,1,1.2,1,0.8,0.4,0.2,0,-0.4,-0.48, 0.6,-0.8,-0.72, 0,0.84,0.99,0.84,0.62,0.26,0,-0.20,-0.40,-0.36,-0.70,-0.58,0,0.59,0.78,0.56,0.30,0,-0.21,-0.24,-0.38,-0.38 0,0.26,0.50,0.39,-0.03,-0.10,-0.12, | 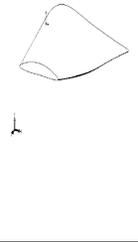 | 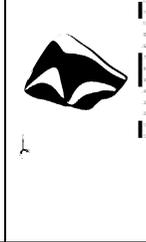 | O1=c O2=c |
| 2 | 0,1.1,1.21,.9,0.82,0.42,0.18,.1,-0.41,-0.46,-0.62,-0.81,-0.70, 0,0.86,0.1,0.82,0.60,0.25,0.01,-0.20,-0.39,-0.39,-0.70,-0.58, 0,0.58,0.76,0.57,0.32,0,-0.21,-0.23,-0.37,-0.39 0,0.26,0.54,0.40,-0.03,-0.1,-0.1, | 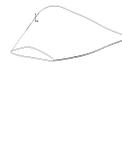 | 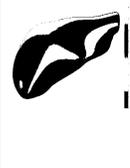 | O1=b O2=c |
| 3 | 0,1,1.2,1,0.8,0.4,0.2,0,-0.4,-0.48,-0.6,-0.8,-0.72, 0,.88,0.99,0.84,0.62,0.26,0,-0.23,-0.35,-0.37,-0.70,-0.54, 0,0.58,0.76,0.58,0.31,0,-0.23,-0.23,-0.37,-0.37 0,0.24,0.50,0.40,-0.03,-0.13,-0.10, | 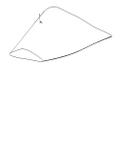 | 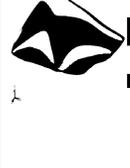 | O1=a O2=e |
| 4 | 0,1.3,1.23,1.06,0.83,0.41,0.28,0.07,-0.41,-0.48,-0.6,-0.8,0.78,0,0.84,.92,0.84,0.62,0.26,0,-0.23,-0.39,-0.37,-0.70,-0.54,0,0.58,0.76,0.58,0.31,0,-0.24,-0.22,-0.36,-0.38, 0,0.24,0.52,0.38,-0.02,-0.12,-0.12, | 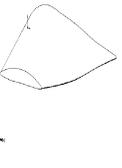 | 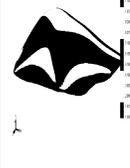 | O1=d O2=c |
| 5 | 0,1.01,1.21,1,0.8,0.4,0.21,0,-0.41,-0.47,-0.59,-0.79,-0.69, 0,0.80,1.01,0.86,0.64,0.26,-0.01,-0.20,-0.40,-0.40,-0.72,-0.56, 0,0.58,0.76,0.58,0.31,0,-0.23,-0.23,-0.37,-0.37 0,0.24,0.52,0.38,-0.06,-0.10,-0.10, | 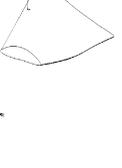 | 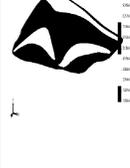 | O1=e O2=d |

In the given example the *target categorical variable* is the value of *Displacements Distribution* calculated by numerical simulation in ANSYS and classified in 4 classes of *a, b, c* and, *d*. In the data sets of geometrical and numerical analysis objective values taken for analysis are given in table.1. This table has gathered initial data sets including the geometry of shapes and numerical simulations from five calculations, based on random initial values of variables.

6.1 Results

The obtained results from preprocessing are available in table.2. Eight variables out of forty two have been selected that have more effects on O1 and, seven variables that have more effects on O2. Two types of classification error (MAE, RMSE) are shown for utilized algorithm corresponding to different classes of objectives. Experiments show that the obtained results are not very sensitive to the exact choice of these thresholds.

Table 2: Variables importance ranking for three classification methods

| Classification Algorithm | MAE | RMSE | Effective Variables | Objectives |
|---|---|---|---|---|
| LADTree | 0.370 | 0.517 | 38,15,24,2,32,41,39,3 | $O_1$ |
| | 0.412 | 0.519 | 41,35,9,17,11,38,37 | $O_2$ |

## 7. Discussion

The whole preprocessing was done within 6.3 minutes on a Pentium IV 2.4 MHZ Processor. The variables were reduced by 50%.
The data set of the given MOO problem was preprocessed and effective variables have been identified. With the results of the preprocessing analyzes presented in table.2 the optimization problem has been much clear in terms of variable and objective interactions. The new created design space based on the new sets of variables listed in table.2 is much smaller which would make the further Meta-modeling based MOO much easier in terms of complexity.  By adjusting the MAE and RMSE in each classification preprocessing the expected number of variables could be arranged. For the given example we were expecting more than a 50% reduction in design space for the errors available in table.2.

## 8. Conclusions

The classification task of data mining has been introduced as an effective option for identifying the most effective variables of the MOO in MCDM systems. The Classification algorithm of LADTree was utilized analyzing the effect of each design variable to the indentified objectives. The number of the optimization variables has been managed very effectively and reduced in the given example.

The modified methodology is demonstrated successfully in the framework. The author believes that the process is simple and fast. Variables were reduced and organized utilizing classification algorithms. The achieved preprocessing results as reduced variables will speed up the process of optimization due to delivered smaller design space and minimum requested computational cost for MOO process. Data mining tools have been found to be effective in this regard. It is evident that the growing complexity of MCDM systems could be handled by a preprocessing step utilizing data mining classification tools.

For future work, studying the effectiveness of the introduced data reduction process in different applications is suggested. Also trying to use other tools of data mining such as clustering, association rules, and comparison could produce beneficial results.

**Acknowledgments**

Author would like to thank his supervisors; Professor Nagy Péter Tibor, Professor Miklós Hoffmann and Daniel Haitas. Also the effort of SIGMA technology to support our academic research is aknowledged.

**A. Mosavi** is a PhD candidate at University of Debrecen; Faculty of Informatics. He received his MS (2007) in automotive engineering from London Kinston University, UK. His current research interest is Multiple Criteria Decision-Making in automotive and aerospace engineering. He has many publications in this subject.